# Auto-Evaluation: A Critical Measure in Driving Improvements in Quality and Safety of AI-Generated Lesson Resources

**Hannah-Beth Clark[1] Margaux Dowland[1] Laura Benton[1] Reka Budai[1] Ibrahim Kaan Keskin[1] Emma Searle[1] Matthew Gregory[1] Mark Hodierne[1] William Gayne[1] John Roberts[1]**

[1]**Oak National Academy**

## ABSTRACT

Designing AI tools for use in educational settings presents distinct challenges; the need for accuracy is heightened, safety is imperative and pedagogical rigor is crucial.

As a publicly funded body in the UK, Oak National Academy is in a unique position to innovate within this field as we have a comprehensive curriculum of approximately 13,000 open education resources (OER) for all National Curriculum subjects, designed and quality-assured by expert, human teachers. This has provided the corpus of content needed for building a high-quality AI-powered lesson planning tool, Aila, that is free to use and, therefore, accessible to all teachers across the country. Furthermore, using our evidence-informed curriculum principles, we have codified and exemplified each component of lesson design. To assess the quality of lessons produced by Aila at scale, we have developed an AI-powered auto-evaluation agent, facilitating informed improvements to enhance output quality. Through comparisons between human and auto-evaluations, we have begun to refine this agent further to increase its accuracy, measured by its alignment with an expert human evaluator. In this paper we present this iterative evaluation process through an illustrative case study focused on one quality benchmark - the level of challenge within multiple-choice quizzes. We also explore the contribution that this may make to similar projects and the wider sector.

**Author Note:** Correspondence concerning this article should be addressed to Hannah-Beth Clark. Email: hannah-beth.clark@thenational.academy

Following the launch of GPT-3.5 in 2022 the edtech market has been flooded with AI tools to support teachers with time-consuming tasks such as lesson planning or generating lesson resources, resulting in a sharp increase in the number of teachers using AI (e.g. Teacher Tapp, 2024). However, a recent Education Endowment Foundation (EEF) research trial highlighted that there is still "limited robust evidence on the use of GenAI for teaching and learning, including the potential impact on disadvantaged pupils, whilst guidance for implementation in schools is also limited" (EEF, 2024).

High-quality teaching is the best lever for improving pupil attainment (UNESCO, 2014; EEF, n.d.) and so it is important that AI tools support teachers in delivering high-quality lessons. Specifically, content created within these tools must be accurate, safe, have research-backed pedagogical design and be specific to the learning and educational context (Baidoo-Anu & Ansah, 2023; UNESCO, 2024). Accuracy is particularly important in this context as inaccuracies and biases in content created by AI tools can exacerbate misconceptions within classrooms and therefore have a detrimental effect on pupil outcomes (Kamalov et al., 2023; Levonian & Henkel, 2024; Murphy & Alexander, 2013; Grassini, 2023). However, there is currently a lack of evaluation accompanying these tools to help assess the quality of content that they are producing (Chiu et al., 2023; EEF,

n.d.), consequently restricting the ability of teachers and organisations to make good decisions about the particular AI tools they use within schools.

As a publicly funded body in the UK, with the aim of improving pupil outcomes and closing the disadvantage gap, Oak National Academy are in a unique position to innovate within this field. We have created a large corpus of 13,000 Open Education Resources (OER) aligned with the national curriculum for England including; slide decks, worksheets, quizzes and videos with transcripts (https://www.thenational.academy/teachers) - designed and quality-assured by expert, subject specialist teachers in line with Oak's evidence-informed curriculum principles (McCrea, 2023). This content is openly licensed, on the Open Government Licence version 3.0 (OGL) which is compatible with Creative Commons by Attribution 4.0 (CC-BY) in line with UNESCO's Recommendation on OER (UNESCO, 2019).

This corpus of high-quality curricula content gives a valuable starting point for an AI-powered lesson planning tool that is free to use and accessible to UK teachers, as research has shown that providing generative AI models with a high-quality corpus in a retrieval database for use in retrieval augmented generation (RAG) can improve accuracy from 67% to 92% (Government Social Research, 2024). In this paper, we describe our approach to designing Aila, our AI lesson assistant and the auto-evaluation agent built alongside to assess the accuracy, quality and safety of the lessons Aila produces. We also present empirical data from a case study to assess the effectiveness of this auto-evaluation agent.

## System Design

Aila is designed to emulate the thought process of an experienced teacher as they plan a lesson. It is intentionally designed not to be a 'single-shot' tool that creates a lesson in one click, but instead supports teacher agency through enabling them to adapt and edit the lesson step-by-step to better suit their students (Figure 1).

Figure 1: Screenshot of Aila's Lesson Planning Interface

Our underlying content, alongside the codification of good practice in lesson design, has enabled us to use several techniques to raise the quality of Aila's outputs. These include RAG, to provide relevant context for the output (Chen et al., 2024) and more specifically content anchoring, to improve lesson quality by instructing the model to respond within the bounds of specified content (i.e. an existing Oak lesson) (Kommineni et al., 2024); prompt engineering, to focus the response of the underlying Large Language Model (LLM) according to our codified definition of a high-quality lesson; and decision-making by the teacher at a granular level to act as the human in the loop (Tsiakas & Murray-Rust, 2022; Wu et al., 2022).

To enable us to understand the effectiveness of these techniques by evaluating Aila's outputs quickly and efficiently, we have built an auto-evaluation agent, using LLM as a Judge methodology (Chiang & Lee, 2023), which is based on Oak's curriculum principles (McCrea, 2023). Each lesson is currently evaluated using a series of auto-evaluation prompts, assessing 24 quality and accuracy benchmarks, such as cultural bias, minimally different quiz answers or the progression of quiz difficulty (for the full list, see Appendix 1). This has enabled us to evaluate the impact of the changes we make to improve Aila and compare the results, such as using different models as the underlying LLM, testing new versions of Aila before release, and identifying particular areas for development, which is the focus of this paper. To validate the tool's effectiveness, we conducted comparative tests between expert human evaluations and auto-evaluations, focused on these identified development areas. While we performed various quality assurance tests, this paper focuses on a specific case study examining the quality of multiple-choice questions (MCQs) in quizzes, one of Aila's most widely used features.

# Case Study

## Method

Aila produces diverse educational resources, including lesson plans and classroom materials. We wanted to understand how closely aligned the auto-evaluation agent was with qualified teachers. To do this we first created a dataset of 2249 user-created Aila lessons, and 2736 lessons produced by Aila without user input or content anchoring (i.e. single shot), totalling 4985 lessons. The lessons were across all four key stages (i.e. for ages 5-16 years) and included maths, English, history, geography and science. The auto-evaluation model (gpt-4o-2024-08-06, temperature: 0.5) scored the lessons on 19 Likert criteria (using a 1-5 scale, see Figure 2) and 5 boolean criteria (true or false), each with their respective justifications.

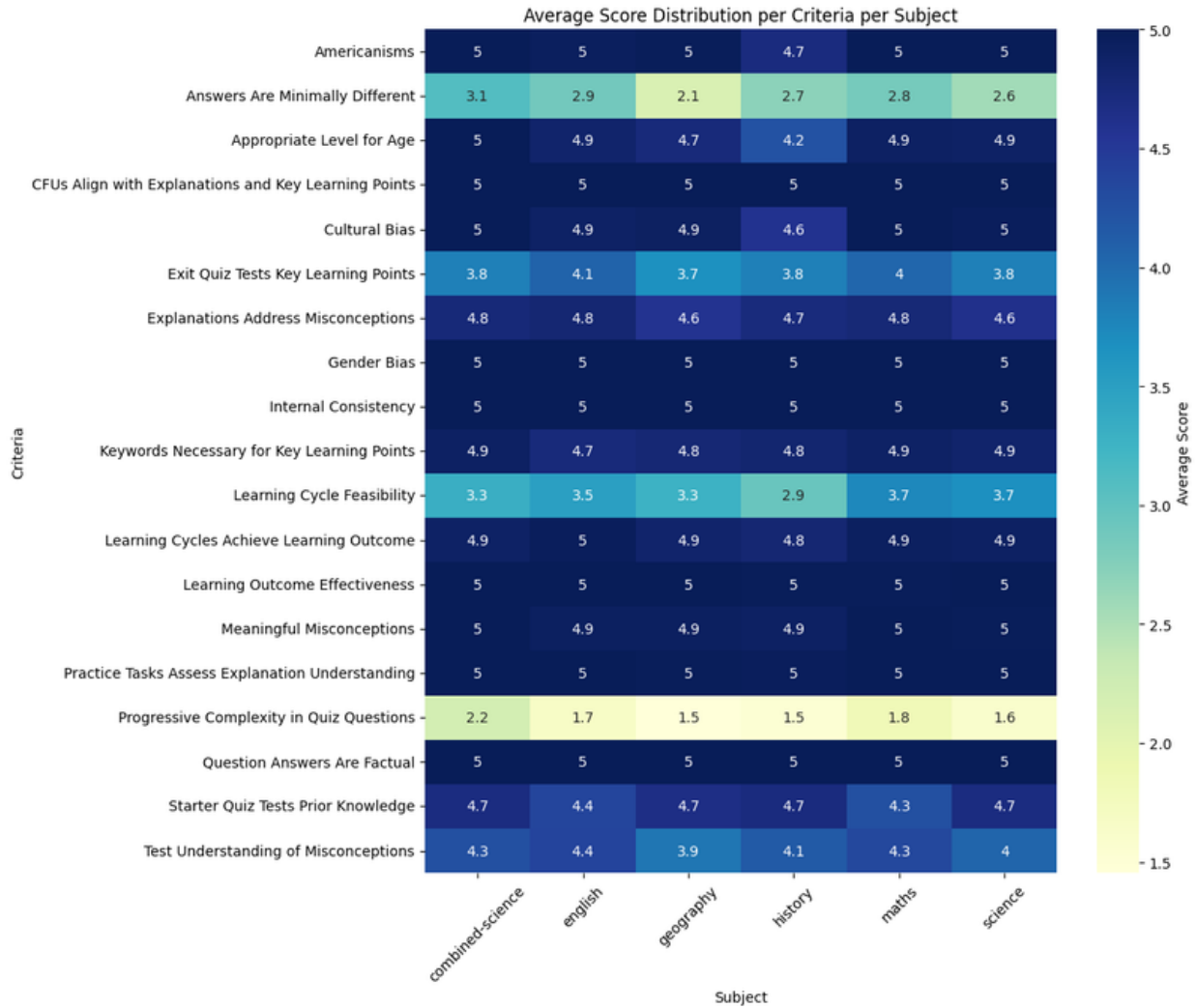

Figure 2: The Auto-Evaluation Tool Assesses Lessons Based on 19 Score-Based Criteria (Excluding 5 Boolean Criteria)

MCQs are particularly valued for their effectiveness in informal assessment and ease of administration in classroom settings (D'Sa & Wisbal-Dionaldo, 20217). Our analysis centred on the quality of MCQ distractors—incorrect but plausible answer choices—as this benchmark consistently scored below average in auto-evaluations. Specifically, we examined the criterion of 'minimally different answers' (see row 2, Figure 2), which evaluates whether distractors are sufficiently subtle to challenge students while effectively assessing concept mastery. Through this work, we wanted to understand two key things: (1) what makes a distractor high or low-quality in providing an appropriate challenge level, (2) how closely aligned the auto-evaluation tool was with qualified teachers.

We recruited 20 qualified teachers (both primary and secondary) who currently work at Oak National Academy, with an average of 14 years of teaching experience, to participate as human evaluators. We asked participants to evaluate randomly assigned MCQs from their subject specialism using the same evaluation criteria as the auto-evaluation tool (see Figure 3). Participants were asked to rate the answers on a scale of 1-5 in terms of the answers being minimally different from each other (with 1 being significantly different and 5 being minimally different) and to add a justification for their score. We found one participant's scores to be unreliable due to us not differentiating between primary and secondary expertise when assigning questions, and these results were subsequently excluded from the analysis. In total, participants assessed 311 multiple-choice questions from the dataset, with an average of 16.4 questions per participant.

## Analysis

Our initial analysis focused on MCQs that teachers scored as 1, 3, and 5 to understand weak, average, and strong distractor quality, conducting a thematic analysis of the teachers' justifications for these scores. We limited our thematic analysis to these three categories to provide clear benchmarks for quality assessment and to identify distinctive characteristics at each level of performance. We then identified exemplar MCQs to supplement the amended auto-evaluation prompts. We also conducted thematic analysis on MCQs where significant discrepancies existed between the auto-evaluation tool and human evaluators (e.g., where the tool scored the quiz 1 or 2 and the teacher gave a 4 or 5) to identify potential blindspots in the automated assessment and understand the reasoning patterns that led to such divergent evaluations.

Finally, we utilised insights gained from the previous two steps to refine the prompt used by the auto-evaluation agent and added few-shot examples (i.e. providing the model with example inputs and outputs). This refinement aimed to incorporate expert teacher feedback into the prompt engineering strategy to better align the auto-evaluation agent's judgments with expert human opinions. We then compared the extent of agreement between expert humans and the auto-evaluation agent both before and after these modifications to assess improvement in alignment.

## Results

### (1) What makes a generated distractor high or low-quality in relation to providing an appropriate level of challenge?

Appendix 2 summarises the key rating justification themes given by the human evaluators. The most common reason for distractors being low-quality was having the *opposite sentiment* to the correct answer (e.g. correct answer is a positive trait and the distractors are all negative traits). Other reasons included having a *different grammatical structure* to the correct answer, as well as the correct answer *repeating words from the question*, but the distractors not. For distractors to be high-quality they should fall into the *same category* as the correct answer, relate to a *common theme*, include *common misconceptions* and have a *similar grammatical structure*.

## (2) How well aligned were the auto-evaluation agent and the human evaluators?

Figure 3 highlights how the auto-evaluation agent was applying excessively strict criteria compared to the human evaluator, rating a large number of quiz questions as having low-quality distractors. It justified the low scores by claiming that the answer options were conceptually very different, thereby lacking the necessary challenge for the specified key stage. There was also an overemphasis on what was expected of students at the key stages, challenging deeper understanding.

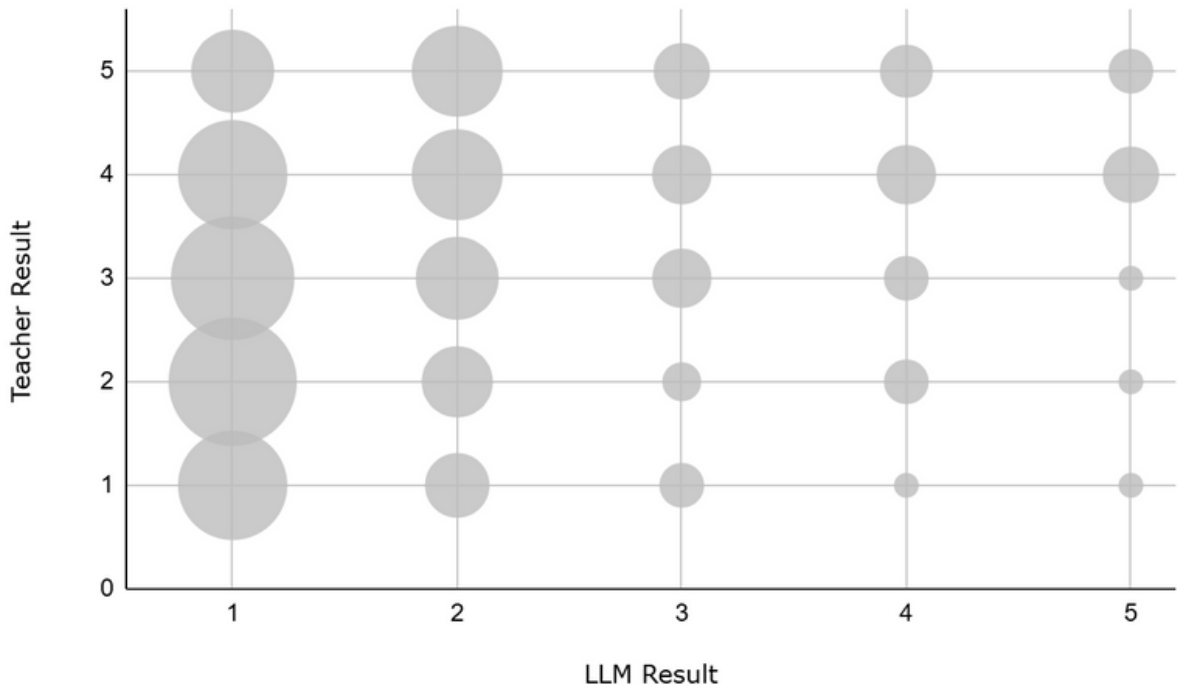

Figure 3: Paired scores of auto-evaluation and human evaluation, bubble size indicates the number of quiz questions. n=311 quiz questions

We used the thematic analysis findings to update the prompt with additional guidance defining a high-quality distractor, and as a result, the auto-evaluation scores and human evaluation scores became more aligned (see Table 1). We calculated the Mean Squared Error (MSE) using the mean of the 10 scores given by the auto-evaluation per evaluation. The mean-based MSE decreased from 3.83 to 2.95 (p-value = 0.00640), which is statistically significant ($p < 0.05$). We also calculated several other evaluation metrics, including the Quadratic Weighted Kappa (QWK), which showed an increase from 0.17 to 0.32, indicating a moderate to large and statistically significant improvement in agreement (see Appendix 3). The justifications given by the LLM now align with the themes identified within the thematic analysis, focusing on plausibility, structural coherence, and commonality (for a detailed example, see Appendix 4).

**Table 1: Count of differences between LLM and human scores before and after improving the prompt based on the thematic analysis findings.**

LLM scores used the mean rounded up to the nearest integer.

| LLM-Human score difference | Count (before) | Percentage (before) | Count (after improving) | Percentage (after improving) |
| --- | --- | --- | --- | --- |
| 0 | 60 | 19% | 85 | 27% |
| 1 | 96 | 31% | 104 | 33% |
| 2 | 80 | 26% | 72 | 23% |
| 3 | 58 | 19% | 37 | 12% |
| 4 | 17 | 5% | 13 | 4% |
| **Total lower scores by LLM** | 232 | 75% | 193 | 62% |
| **Total higher scores by LLM** | 19 | 6% | 33 | 11% |

## Discussion

Through an illustrative case study, we have demonstrated the potential of using an auto-evaluation agent to drive improvement in the quality of AI-generated lessons and resources, as well as how the effectiveness of this agent can be improved by drawing on specific teaching expertise of human evaluators. Thematic analysis of rating justifications allowed us to codify what high and low-quality distractors looked like (with few-shot examples) and incorporate this information directly into the prompt, increasing the alignment with the human evaluators and driving improvements in the overall MCQ quality.

Incorporating the thematic analysis and corresponding representative examples for scores of 2 and 4 in future work could help reduce minor discrepancies by increasing granularity, especially in cases where scores are '1 away' from human evaluations. Absolute alignment is not necessarily the ultimate goal; the more important measure of success would be to see if the justifications the LLM gives alongside scores of 1, 3 and 5 are in line with the themes we found, providing consistent scoring according to these guidelines. Further thematic analysis would be required to establish this. Even after the changes, the LLM still scores lower than the human the majority of the time. This greater sensitivity is more beneficial than the alternative, as potential issues are more likely to be flagged and addressed.

There were also limitations to this work. We had a specific focus on answer differentiation and MCQs which could have implications for wider generalisability. Furthermore, due to time constraints, we weren't able to have multiple human evaluators for each question. Ideally, we would have an average human score per evaluation to deal with possible outliers. In future work, we could also consider weighting these responses according to the teacher's experience level, factoring in years of experience, teaching role and other metrics.

## Recommendations

Aila has been designed specifically to support teachers in the UK with planning high-quality lessons and resources to reduce teacher workload and improve the quality of materials produced using AI. We hope by sharing what we have learned through this work it can also have an impact on other projects:

1. Having a base of high-quality OER has been integral to the quality of lessons produced by Aila. Our curriculum materials are aligned with the national curriculum for England, produced by expert teachers, available on an open government licence, and targeted at UK schools. *For other organisations looking to develop tools within this space in other contexts, access to high-quality resources appropriate for their context will be imperative. We seek to enable this by making our OER resources available through a public API.*
2. We had already done significant work codifying and exemplifying high-quality curriculum design. This provided invaluable input as the starting point for writing our prompt and, in turn, our evaluation tools. *Deciding on your organisation's agreed-upon concept of "high-quality" is an important starting point before developing your tool, as this will be built into your prompt and evaluation work.*
3. Using a cycle of comparative auto and human evaluations allowed us to continuously iterate on the auto-evaluation prompt and will ultimately also enable us to refine Aila's prompt. *Once you have identified full lesson plans that achieve good scores aligned between evaluators through this iterative process these plans can subsequently be used to fine-tune generation models to output better quality lesson plans* [(Ouyang et al., 2022)](#).

## Concluding remarks

We believe that auto-evaluation is a powerful tool for driving improvement in AI-produced content quickly and efficiently. We have focused specifically on a "quality" benchmark but we are also in the process of applying this approach to our "safety" benchmarks. The use of our auto-evaluation tool to evaluate different versions of Aila as we release them, comparisons of quality in how RAG is used, and the use of fine-tuning to develop the quality of our AI tools are further areas we plan to investigate. We also aim to use an improvement agent which will take feedback from our auto-evaluation agent to improve the quality of lesson content before it is displayed to users as well as suggest specific areas for users to check carefully or improve.

All our OER content is being made available on an OpenAPI ([https://open-api.thenational.academy/](https://open-api.thenational.academy/)), and our code - including the prompts - is openly licenced ([https://github.com/oaknational/oak-ai-autoeval-tools](https://github.com/oaknational/oak-ai-autoeval-tools)),

allowing others innovating within this field to use it as a starting point (for examples of our current evaluation prompts, see Appendix 5). We therefore hope that this work will provide a framework and useful starting point for other organisations to develop similar approaches to enable them to objectively evaluate their content, driving forward progress within their own tools.

## Acknowledgements

We are grateful to all of the Oak National Academy teachers who contributed their valuable expertise to this work as well as the wider Aila squad for supporting the development and improvement of the tool. We would also like to thank Owen Henkel and Manolis Mavrikis for their helpful suggestions in improving this paper.

In lieu of an IRB, internal research participation policies were implemented and all participants were provided with an information sheet and completed an online consent form to ensure the research was ethically sound (see Appendix 6).

## Appendices

### Appendix 1

Full set of assessed quality and accuracy benchmarks:

| Prompt Criteria Title | Check Output Format | Relevant Lesson Plan Part 'lesson' implies entry lesson | Criteria Group |
| --- | --- | --- | --- |
| Learning Cycle Feasibility | Likert | Key-Stage, Cycle-feedback, Cycle-practice, Cycle-explanations, Cycle-check | learning-cycles |
| Practice Tasks Assess Explanation Understanding | Likert | Cycle-practice, Cycle-explanations | learning-cycles |
| Keywords Necessary for Key Learning Points | Likert | Keywords, Key Learning Points | learning-outcomes |
| CFUs Align with Explanations and Key Learning Points | Likert | All cycles | learning-outcomes |
| Learning Cycles Achieve Learning Outcome' | Likert | Learning Outcome, Learning Cycle | learning-outcomes |
| Learning Outcome Effectiveness | Likert | Learning Outcome | learning-outcomes |
| Explanations Address Misconceptions | Likert | Cycle-explanations, Misconceptions | misconceptions |
| Meaningful Misconceptions | Likert | Misconceptions, Topic | misconceptions |
| Test Understanding of Misconceptions | Likert | Exit Quiz, Cycle-check, Misconceptions | misconceptions |
| Question Answers Are Factual | Likert | Whole Lesson | lesson-quality |
| Internal Consistency | Likert | Whole Lesson | lesson-quality |
| Appropriate Level for Age | Likert | Whole Lesson, Key-Stage | bias |
| Americanisms | Likert | Whole Lesson | bias |
| Cultural Bias | Likert | Whole Lesson | bias |
| Gender Bias | Likert | Whole Lesson | bias |
| Exit Quiz Tests Key Learning Points | Likert | Exit Quiz, Key Learning Points | quizzes |
| Starter Quiz Tests Prior Knowledge | Likert | Starter Quiz, Prior Knowledge | quizzes |
| Answers Are Minimally Different | Likert | Starter Quiz, Exit Quiz | quizzes |
| Progressive Complexity in quiz Questions | Likert | Exit Quiz | quizzes |
| Learning Cycles Increase in Challenge | Boolean | Learning Cycles | learning-cycles |
| No Negative Phrasing in Quiz Questions | Boolean | Starter Quiz, Exit Quiz | quizzes |
| Repeated Questions in Quizzes | Boolean | Exit Quiz, Starter Quiz | quizzes |
| Starter Quiz does not Rest Lesson Content | Boolean | Starter Quiz, Learning Cycles, Learning Outcome, Key Learning Points, Prior Knowledge | quizzes |
| Exit Quiz Contains Vocabulary Question | Boolean | Exit Quiz, Keywords | quizzes |

# Appendix 2

Summary of thematic analysis based on 278 quiz questions and answers, including the initial set of quiz questions, excluding scores of 2 and 4

| Mean human score | Theme | Frequency | Example |
|---|---|---|---|
| 1.5 | Distractors have an opposite sentiment to the right answer or the question | 27 | Which of these is a positive impact of TNCs in the food industry? **Creating jobs in developing countries,** *Making local foods more expensive., Eliminating smaller companies., Reducing dietary variety* |
| 1.9 | The correct answer is structurally different | 21 | What is the Atacama Desert known for? **Being one of the driest places on Earth**, *Its large rainforest, Its snowy mountains, Its tropical beaches* |
| 1.7 | The right answer repeats words from the question | 11 | What is required to simplify an algebraic fraction by factorisation? **Factorise the quadratic expressions in the numerator and denominator.** *Multiply the numerator and denominator by a common factor., Add the expressions in the numerator and denominator., Subtract the denominator from the numerator.* |
| 3.0 | One of the distractors in semantically different | 21 | Which period did William Wordsworth belong to? **Romantic** *Victorian, Elizabethan, Modernist* |

| 3.0 | No obvious mistakes in the quiz but it lacks sufficient challenge | 17 | What does TNC stand for? **Transnational Corporation**, *Total National Company*, *Trade Negotiation Committee*, *Territorial Network Corporation* |
| --- | --- | --- | --- |
| 2.6 | Distractors do not address typical misconceptions | 10 | If the probability of an event is 0.5, what is the probability of the opposite event? ***0.5***, *1*, *0*, *It cannot be determined* |
| 3.0 | Two options are semantically very different to the other two options | 6 | What is an astronaut? ***A scientist trained to go into space,*** *A space tour guide*, *A pilot who flies airplanes*, *A doctor who treats illnesses* |
| 3.0 | Answer options fall into different categories | 4 | Which term describes medieval stories of knights and romance? ***Chivalry***, *Allegory*, *Satire*, *Fable* |
| 5.0 | All quiz answers fall into the same category | 14 | Which material is not commonly used to make shell structures? ***Wood***, *Cardboard*, *Plastic*, *Metal* |
| 5.0 | All answers relate to a common theme | 10 | Which term describes the complete outer edge of a circle? ***circumference***, *radius*, *diameter*, *segment* |
| 5.0 | Distractors include common misconceptions | 8 | What was the primary crop grown on plantations in early Virginia? ***Tobacco***, *Cotton*, *Sugar*, *Rice* |

| 5.0 | All answers are structurally similar | 4 | How does Wordsworth portray nature in the poem? *As beautiful and restorative,* As mundane and uninteresting, As artificial and man-made, As oppressive and confining |

## Appendix 3

*Evaluation metrics for each score category before and after improving the prompt based on the thematic analysis findings, using the mean of the LLM scores rounded to the nearest integer.*

| Results before the thematic analysis | | | | | |
| --- | --- | --- | --- | --- | --- |
| Score | Precision | Recall | F1 | Accuracy | QWK |
| 1 | 0.19 | 0.78 | 0.30 | 0.19 | 0.17 |
| 2 | 0.14 | 0.16 | 0.15 | - | - |
| 3 | 0.26 | 0.07 | 0.11 | - | - |
| 4 | 0.46 | 0.08 | 0.14 | - | - |
| 5 | 0.29 | 0.04 | 0.07 | - | - |

| Results after the thematic analysis | | | | | |
| --- | --- | --- | --- | --- | --- |
| Score | Precision | Recall | F1 | Accuracy | QWK |
| 1 | 0.22 | 0.78 | 0.35 | 0.27 | 0.32 |
| 2 | 0.24 | 0.18 | 0.20 | - | - |
| 3 | 0.24 | 0.11 | 0.15 | - | - |
| 4 | 0.43 | 0.27 | 0.33 | - | - |
| 5 | 0.50 | 0.18 | 0.26 | - | - |

## Appendix 4

**Question:** What type of government did France become immediately after the Revolution?

**Answer:** A constitutional monarchy

**Distractors:** A democratic republic, An absolute monarchy, A socialist state

**Teacher Score:** 4

**Teacher justification:** valid, and high-quality, distractors, although we may not expect pupils to have an understanding of socialism when studying the french revolution, making 4 not as strong.

**Original Prompt Score:** 2

**Original Prompt Justification:** The question asks about the type of government France became immediately after the Revolution, with the correct answer being 'A constitutional monarchy.' The distractors include 'A democratic republic,' 'An absolute monarchy,' and 'A socialist state.' These options are distinct in terms of political systems and ideologies. While they all represent different forms of governance, they are not minimally distinct from each other, as each represents a fundamentally different political structure. The correct answer requires specific historical knowledge rather than nuanced understanding of similar concepts. Given the Key Stage 3 level, the question is appropriate in complexity but does not challenge deeper discernment skills due to the clear distinctions between the options.

**Mean Score Over 10 Runs:** 1.5

**Improved Prompt Score:** 4

**Improved Prompt Justification:** The distractors are plausible as they represent different forms of government that could have been considered after a revolution. However, 'A socialist state' is less plausible for the immediate aftermath of the French Revolution, as socialism was not a prominent political ideology at that time. The distractors share a commonality with the correct answer as they all describe types of government. Structurally, all options are similar in format. Overall, the distractors are not completely distinct but could be more refined to better test understanding.

**Mean Score Over 10 Runs:** 3.7

# Appendix 5

*The evaluation objective the teacher was presented with:*

Evaluate whether the answers and distractors are minimally different from each other. This assessment should determine if the quiz questions are designed to subtly test the students' deeper understanding of the topic by making the correct answers only slightly more accurate than the distractors. Ensure that the minimal differences are still sufficient to test knowledge accurately without being trivial or overly ambiguous.

**The original prompt:**

Objective:

You will be given a quiz question from a lesson plan as well as the key stage of the lesson. Your task is to evaluate the question based on the following criteria:

Evaluate whether the answers and distractors are minimally different from each other. This assessment should determine if the quiz questions are designed to subtly test the students' deeper understanding of the topic by making the correct answers only slightly more accurate than the distractors. Ensure that the minimal differences are still sufficient to test knowledge accurately without being trivial or overly ambiguous. Take into account the key stage of the lesson and the complexity of the topic when assessing the question.

Note: A thoughtful analysis of the question is required. Submissions that do not demonstrate a detailed examination will be disregarded.

Question:

{{question}}

(End of Question)

Key Stage:

{{key_stage}}

(End of Key Stage)

Rating Criteria:

5 (Minimally Distinct): The correct answers and distractors are subtly different, effectively assessing deeper understanding and discernment skills without being obvious, while still ensuring each question is valid and distinct enough to be fair.

1 (Clearly Distinct): The correct answers and distractors are clearly and easily distinguishable, which may not effectively challenge the student's deeper understanding or discernment skills.

Provide Your Rating:

Rate the quiz question on a scale from 1 to 5 based on how minimally distinct the answers and distractors are. A score of 5 means the differences are minimal and require careful thought to discern, yet are meaningful and fair, while a score of 1 indicates the options are too obvious or too similar, potentially undermining the test of deeper understanding.

Format your response according to the JSON structure below, providing a justification for your score. Your justification should be concise, precise and directly support your rating.

Use this JSON format for your evaluation:

{

"justification": "<JUSTIFICATION>",

"result": "<SCORE>"

}

Only answer with the format above and return a single score, not a collection of scores.

A detailed justification is crucial, even for a score of 5.

**The improved prompt:**

Objective:

You will be given a quiz question from a lesson plan as well as the key stage of the lesson. Your task is to evaluate the question based on the following criteria:

Evaluate whether the answers and distractors are minimally distinct from each other (on a scale of 1-5, with 1 being the lowest). The criteria below are aspects to consider when making your evaluation, but failing to meet one of these points does not automatically result in a low score. They are simply factors that may influence the rating.

Criteria to look for:

Plausibility:

- Consider whether the distractors (incorrect answers) are plausible enough to seem like potential correct answers
- Do any of the distractors stand out as obviously incorrect or unrelated to the question?
- Are the distractors plausible due to the fact that they test common misconceptions, making them viable options for the students?
- Ensure that distractors are plausible enough to test understanding, even if conceptually different

Sample Inputs and Outputs for Plausibility:

Input: {'answers': ['Tobacco'], 'question': 'What was the primary crop grown on plantations in early Virginia?', 'distractors': ['Cotton', 'Sugar', 'Rice']}

Output: {"justification": "The first distractor is a clear plausible distractor, and the other two would highlight geographical misconceptions.", "result": "5"}

Input: {'answers': ['0.5'], 'question': 'If the probability of an event is 0.5, what is the probability of the opposite event?, 'distractors': ['1', '0', 'it cannot be determined']}

Output: {"justification": "The distractors would be better if they included 0.2 or -0.5 to address common misconceptions", "result": "3"}

Input: {'answers': ['Transnational Corporation'], 'question': 'What does TNC stand for?', 'distractors': ['Total National Company', 'Trade Negotiation Committee', 'Territorial Network Corporation']}

Output: {"justification": "The distractors need to be closer to the correct answer e.g. all ending in corporation/starting with trans", "result": "3"}

Input: {'answers': ['Dancing'], 'question': 'In 'I Wandered Lonely as a Cloud', the daffodils are personified as:', 'distractors': ['Singing', 'Running', 'Sleeping']}

Output: {"justification": "'Singing' could be changed as 'Wandering' clearly pertains to movement. 'Sleeping' could also be stronger.", "result": "3"}

---

Commonality:

- Evaluate whether the distractors share something in common with the correct answer.
- Do the distractors fit within the same category or concept as the correct answer, even if they differ in specific details?
- Even if distractors differ in some aspects, ensure they still relate to the concept being tested

Sample Inputs and Outputs for Commonality:

Input:

- Answers: ['Wood']
- Question: Which material is not commonly used to make shell structures?
- Distractors: ['Cardboard', 'Plastic', 'Metal']

Output:

- Justification: "All of the answers are materials used to build structures so could all reasonably be correct."
- Result: "5"

Input:

- Answers: ['Chivalry']
- Question: Which term describes medieval stories of knights and romance?
- Distractors: ['Allegory', 'Satire', 'Fable']

Output:

- Justification: "These distractors are broadly plausible because they are difficult literary terms, but they are not great distractors because they are not similar to the correct answer."
- Result: "3"

Input:

- Answers: ['Romantic']
- Question: Which period did William Wordsworth belong to?
- Distractors: ['Victorian', 'Elizabethan', 'Modernist']

Output:

- Justification: "Modernist is a bit of an outlier as it is very much 20th Century."
- Result: "3"

---

Structural Coherence:

- Assess whether the distractors and correct answer are structurally similar (e.g., similar in length, format, or complexity), making it easier to guess the correct answer.
- Does the correct answer repeat keywords or phrases from the question?
- Do any distractors have the opposite sentiment to the question, making them easy to rule out?
- Differences in structure are acceptable if they contribute to assessing understanding

Sample Inputs and Outputs for Structural Coherence:

Input:

- Answers: ['As beautiful and restorative']
- Question: How does Wordsworth portray nature in the poem?

- Distractors: ['As mundane and uninteresting', 'As artificial and man-made', 'As oppressive and confining']

Output:

- Justification: "These distractors are strong as they all contain the same format of 'As *** and ****' and relate to one another."
- Result: "5"

Input:

- Answers: ['Creating jobs in developing countries.']
- Question: Which of these is a positive impact of TNCs in the food industry?
- Distractors: ['Making local foods more expensive.', 'Eliminating smaller companies.', 'Reducing dietary variety.']

Output:

- Justification: "The correct answer is the only one positive answer listed."
- Result: "1"

Input:

- Answers: ['Being one of the driest places on Earth']
- Question: What is the Atacama Desert known for?
- Distractors: ['Its large rainforest', 'Its snowy mountains', 'Its tropical beaches']

Output:

- Justification: "The distractors have a different format to the correct answer."
- Result: "1"

Input:

- Answers: ['Factorise the quadratic expressions in the numerator and denominator.']
- Question: What is required to simplify an algebraic fraction by factorisation?
- Distractors: ['Multiply the numerator and denominator by a common factor.', 'Add the expressions in the numerator and denominator.', 'Subtract the denominator from the numerator.']

Output:

- Justification: "The correct answer is the only one with 'factorise' in it."
- Result: "1"

---

Note: A thoughtful analysis of the question is required. Submissions that do not demonstrate a detailed examination will be disregarded.

Question:

{{question}}

(End of Question)

Key Stage:

{{key_stage}}

(End of Key Stage)

Rating Criteria:

5: Minimally distinct

1: Clearly distinct

Provide Your Rating:

Rate the quiz question on a scale from 1 to 5 based on how minimally distinct the answers and distractors are.

For cases where the distractors partially meet the criteria but are not fully aligned with the examples for scores of 1, 3, or 5, assign scores of 2 or 4 based on how closely they match the higher or lower examples. If a theme arises outside of those listed, evaluate it based on the overall quality and coherence of the answers and apply the most appropriate score.

Format your response according to the JSON structure below, providing a justification for your score. Your justification should be concise, precise and directly support your rating.

Use this JSON format for your evaluation:

{

"justification": "<JUSTIFICATION>",

"result": "<SCORE>"

}

Only answer with the format above and return a single score, not a collection of scores.

A detailed justification is crucial, even for a score of 5.

# Appendix 6

**Participant information sheet**

We have built an auto-evaluation tool to enable us to evaluate the quality of content produced by Aila quickly, enabling us to test changes before we release them. This will allow us to rapidly iterate and improve Aila - our AI Lesson Assistant, in an efficient and resource-effective way.

**Who can take part?**
We are looking for qualified teachers who have a good understanding of their subject and key stage pedagogy to participate. If you are a primary teacher, you will be evaluating content produced for KS1 and 2 lessons across a range of subjects. If you are a secondary school teacher, you will evaluate content for just your specialist subject. You must have a Google email address to take part at the moment.

**How long will it take?**

We are adding new evaluations each week. For each evaluation, there are normally about 200 different lessons or sections of lessons available for you to evaluate. Please complete as many as you have time for, although we ask that you do a minimum of 10 lessons (approximately 10 minutes).

**What will I be doing?**

You will be shown an evaluation criteria and lesson/part of a lesson. You will be asked to use the criteria given to evaluate that section of the lesson. You may be asked to give a score on a Likert scale (a number scale) or a Boolean scale (true/false). For example, you may be looking at whether a quiz contains any Americanisms. You will then look at the quiz and select 'true' or 'false' depending on whether there are Americanisms present. You will be asked to give a justification for your answer.

**What should I include in my justification?**

The justification is our way of transferring the knowledge of an expert teacher to our auto-evaluation tool. Imagine you were reviewing a trainee teacher's planning and they had included something brilliant that you wanted to praise, or something that needed improving. Your justification should be the explanation that you would give to them that would help them to understand why their planning was really strong or why it needed to be developed.

**What if I am unsure of what evaluation to give for a question?**

Just click 'skip'. We are relying on our human evaluations to be our point of truth.

**What if I finish all the evaluations?**
You are a hero! We will release a new set of evaluations each week so keep your eyes peeled for the next set!

**What will we do with the results that we collect?**
We are collecting human evaluations to compare to the evaluations that our auto-evaluation tool has completed. Where the auto-evaluation is poorly aligned with the expert human evaluators, we will refine the auto-evaluation tool, using the justifications you have provided, to make the auto-evaluation tool more accurate.

**Are they real lessons that I am evaluating?**

We have created a test set of lessons for this process. These intentionally contain lessons of varied quality so that we can train our auto-evaluation tool to evaluate a full range of lessons and resources effectively.

**What will happen if you log out and come back in?**
Just click 'skip this question' until you get to the last question you submitted.

**What personal data will you collect?**

As part of this, we will collect the following personal data: *Your name and email* so that you can log into the human evaluation tool. We will also collect your *school/organisation* and *job role* to help us better understand the diversity of participants.

**What is the legal basis for collecting this personal data?**
The Data Protection Act 2018 and the UK General Data Protection Regulation (GDPR) state that organisations must not use personal data unless they have a legal basis for doing so. The legal basis that we rely on is "consent" (as set out in Article 6(1)(a) of the GDPR).

This means that we rely on your active consent to process your personal data when you participate in the study. This consent is provided by completing the consent form below (name, email).

**What will happen to my data?**
Your personal data will be treated confidentially and will not be shared with any third party organisation outside of Oak National Academy. The aggregated and anonymised dataset may be shared with third-parties (e.g. universities for use in student projects) and research findings may be shared through external reports, publications and other outputs. Anonymised direct quotes from the written justifications may be used in these outputs.

We will minimise the processing of personal data where possible. Your personal data will be kept on our records for 5 years in line with our Privacy Policy.

It is your choice to take part, but please be aware once you begin the human evaluation we may not be able to delete your data from the research project if we have already begun the data analysis or if the anonymised data set has already been shared.

**I want to take part - what do I need to do next?**

1. Ensure you have read the **participant information sheet** above and completed the below **consent form.**

2. Click **here** to get started.

3. **Sign in** with your Google email and fill in your details *(it will remember you the next time you log in!)*

4. Select the '**start evaluation**' tab on the left.

5. Read the **evaluation criteria** at the top of the page and the **lesson/section of a lesson**.

6. Evaluate the component against the **evaluation criteria** *(you will either give a score or select an option depending on the evaluation criteria you are completing)* and give a **justification** for your score.

**A reminder: you are ONLY evaluating based on the evaluation criteria.**

**7.** Click **submit.**

(Image courtesy of Google DeepMind on Unsplash. License CC-BY.)

↵

- D'Sa, J. L., & Wisbal-Dionaldo, M. L. (2017). Analysis of multiple choice questions: item difficulty, discrimination index and distractor efficiency. *International Journal of Nursing Education*, *9*(3). https://doi.org/10.1016/j.mjafi.2020.11.007

↵

- EEF. (2025). *ChatGPT in lesson preparation - Teacher Choices trial.* Education Endowment Foundation. https://educationendowmentfoundation.org.uk/projects-and-evaluation/projects/choices-in-edtech-using-generative-ai-chatgpt-for-ks3-science-lesson-preparation-2024-teacher-choices-trial

↵

- EEF. (n.d.). *High-quality teaching*. Education Endowment Foundation. Retrieved 8 January, 2025, from https://educationendowmentfoundation.org.uk/support-for-schools/school-planning-support/1-high-quality-teaching

↵

- Government Social Research. (2024, October). *Use Cases for Generative AI in Education: Building a proof of concept for Generative AI feedback and resource generation in education contexts* [Technical report]. GOV.UK. https://assets.publishing.service.gov.uk/media/671108a18a62ffa8df77b2bf/Use_Cases_for_Generative_AI_in_Education_-_Technical_report_October_2024.pdf

↵

- Grassini, S. (2023). Shaping the future of education: exploring the potential and consequences of AI and ChatGPT in educational settings. *Education Sciences*, *13*(7), 692. https://doi.org/10.3390/educsci13070692

↵

- Kamalov, F., Santandreu Calonge, D., & Gurrib, I. (2023). New Era of Artificial Intelligence in Education: Towards a Sustainable Multifaceted Revolution. *Sustainability*, *15*(16), 12451. https://doi.org/10.3390/su151612451

↵

- Kommineni, V. K., König-Ries, B., & Samuek, S. (2024). From human experts to machines: An LLM supported approach to ontology and knowledge graph construction. *arXiv preprint*, *2403.08345*. https://arxiv.org/pdf/2403.08345

↵

- Levonian, Z., & Henkel, O. (2024). *Safe Generative Chats in a WhatsApp Intelligent Tutoring System* [Workshop paper]. Human-Centric eXplainable AI in Education and the Leveraging Large Language